\documentclass[11pt,twoside]{article}

\usepackage{asp2006}
\usepackage{epsf}

\begin{document}
\title{Panchromatic Views of Large-scale Extragalactic Jets}  
\author{C. C. Cheung\altaffilmark{1}} 
\affil{Kavli Institute for Particle Astrophysics and Cosmology\\Stanford
University, Stanford, CA 94305, USA}

\altaffiltext{1}{Jansky Postdoctoral Fellow of the National Radio
Astronomy Observatory. The NRAO is operated by Associated Universities, 
Inc. under a cooperative agreement with the NSF.}

\begin{abstract} Highlights of recent observations of extended jets in AGN are
presented. Specifically, we discuss new spectral constraints enabled by {\it
Spitzer}, studies of the highest-redshift (z$\sim$4) radio/X-ray quasar jets, and a
new {\it VLBA} detection of superluminal motion in the M87 jet associated with a
recent dramatic X-ray outburst.  Expanding on the title, inverse Compton emission
from extended radio lobes is considered and a testable prediction for the gamma-ray
emission in one exemplary example is presented. Prospects for future studies with
{\it ALMA} and low-frequency radio interferometers are briefly described. 
\end{abstract}

\section{Shedding New Light on Large-scale Extragalactic Jets}

{\it Chandra} observations have established that X-ray emission is a common feature
of radio jets in active galactic nuclei (AGN) on kiloparsec scales. Harris \&
Krawczynski (2006, and the accompanying
website\footnote{http://hea-www.harvard.edu/XJET/ lists 78 X-ray jet and hotspot
detections as of Dec 2006.\label{foot1}}) provided an extensive recent review of
this topic.  Here, we focus on some of the more recent observational results
reported subsequent to this review.

\begin{figure*}[hb] \plotone{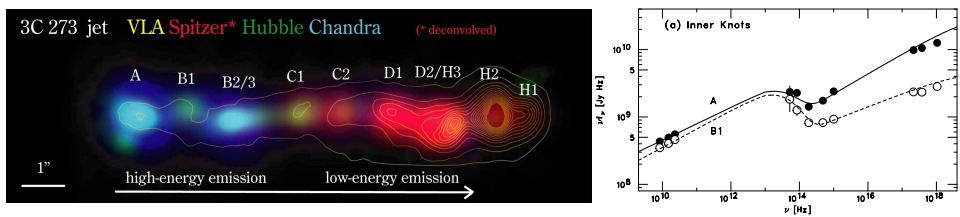} \caption{Multi-band
arcsec-resolution image of the 3C~273 jet and sample knot SEDs (from Uchiyama et
al. 2006).\label{fig-1}}\end{figure*}

Multi-wavelength studies of radio jets have relied on the sub-arcsecond resolution
imaging capabilities of {\it Hubble} and {\it Chandra};  additional spectral
constraints are now being obtained with {\it Spitzer}.  Recent mid-IR detections of
the 3C~273 jet (Uchiyama et al. 2006) and the hotspots in Cygnus A (\L. Stawarz et
al., in preparation) have clarified the relationship between the X-ray and
lower-energy emission. Specifically in 3C~273, most of the optical/UV emission
($\sim$10$^{15}$ Hz) in the inner jet knots lie essentially on a power-law
extrapolation of the X-rays (Fig.~\ref{fig-1}), providing definition of this
high-energy spectral component. Since the jet is long known to be optically
polarized, an indelible signature of synchrotron radiation, it is reasonable to
infer a synchrotron origin for the X-rays also.  The significance of this result to
the wider applicability of synchrotron and inverse Compton (IC off the CMB) models
in other X-ray jets is so far not clear since this spectral signature has not been
defined so clearly in any other extragalactic jet.  New {\it Spitzer} observations
of 10 other powerful X-ray/radio jets selected from XJET$^{\ref{foot1}}$ are being
obtained (PI: Y.  Uchiyama) and optical/UV polarization imaging will be useful in
this respect.

\begin{figure*}[t]
\plotone{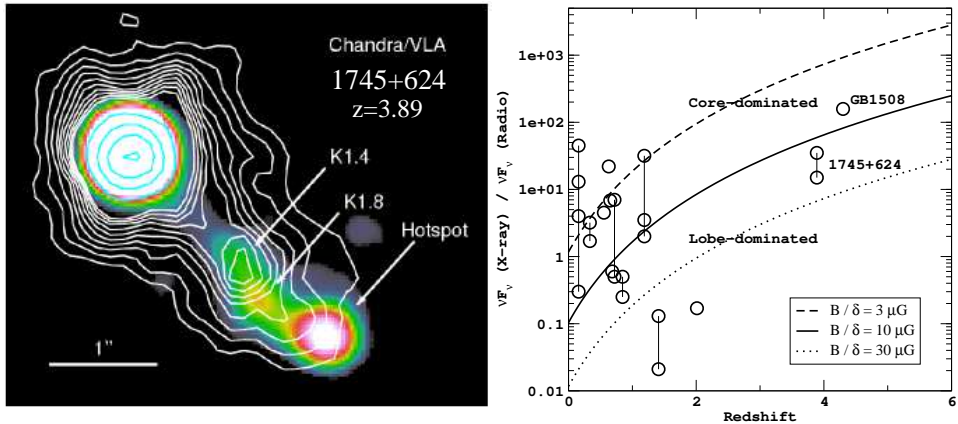}
\caption{[left] The
X-ray (contours) and radio (color) jet in the $z$=3.9 quasar 1745+624
(from Cheung et al. 2006; 1\hbox{$^{\prime\prime}$}=7.2 kpc).
[right] Plot of observed $f_{\rm x}$/$f_{\rm r}$ vs. redshift
for X-ray jets in quasars (adapted from Cheung 2004, with the addition of
the 1745+624 case). Vertical lines
connect different knots from the same jet. Curves indicate expected
$f_{\rm x}$/$f_{\rm r}$ ratio for the given combinations of
$B$ and $\delta$ (for $\alpha$=1), which in the IC/CMB model, scale as
(1+$z$)$^{4}$. The
central solid curve approximately separates knots in core-dominated (larger
$\delta$) and lobe-dominated (smaller $\delta$) sources.} \label{fig-2}
\end{figure*}

One potentially distinguishing feature between synchrotron and IC/CMB models is
that the latter predicts high-$z$ quasars to have bright X-ray jets because of the
strong $z$-dependence of the CMB energy density. Specifically, we expect an
observed monochromatic X-ray to radio flux ($f_{\nu}\equiv\nu$$F_\nu$) ratio,
$f_{\rm x}/f_{\rm r} \simeq u_{\rm cmb}/u_{\rm B} \simeq 10
(1+z)^{4}\delta^{2}/B^2_{\rm \mu G}$ if IC emission dominates ($\delta$ is the
relativistic Doppler factor and $\alpha$=1 is assumed), whereas we expect $f_{\rm
x}/f_{\rm r}\propto z^{0}$ in the synchrotron case. Two very high-redshift
($z$$\sim$4) jets are currently clearly detected in both X-ray and radio bands
(Cheung, Stawarz, \& Siemiginowska 2006, and refs. therein).  The observed $f_{\rm
x}/f_{\rm r}$ ratio of these jets are systematically larger than in lower-$z$
examples (Fig.~\ref{fig-2}), as one expects in the IC/CMB scenario. Differences in
$B$ and jet $\delta$ then accounts for the widely varying $f_{\rm x}/f_{\rm r}$
ratios in different jets at similar redshifts;  this picture makes clear
predictions for jets aligned away from our line of sight, i.e., lobe-dominated
quasars. {\it Chandra} observations of four additional $z$$\sim$4--5 quasars with
known arcsec-scale radio jets being obtained in the current cycle will shed further
light on this issue.

\begin{figure*}[ht]
\plotone{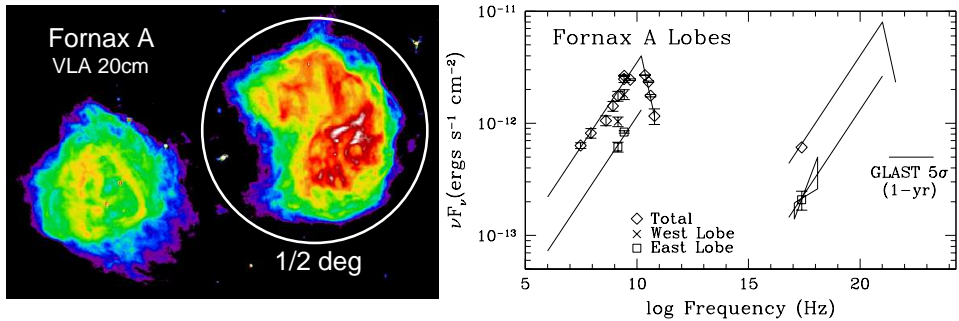}
%\plotfiddle{Fornax.ps}{1.4in}{-90}{36}{36}{-143}{163.0}
%First val: 1.8in, moves figure down
\caption{Radio image of the double-lobed radio galaxy Fornax A (data from Fomalont
et al. 1989) and SEDs of its different components. At a distance of 18.6 Mpc, the
0.5\hbox{$^\circ$} diameter circle corresponds to 162 kpc. The low-frequency radio
data were compiled by Isobe et al. (2006; Fig.~5 therein) with additional 23--61
GHz data from {\it WMAP} (the 94 GHz measurement is omitted because of its lower
significance). The predicted IC/CMB components assume $B$=1.5$\mu$G in the lobes to
match the detected X-rays.} \label{fig-3} \end{figure*}

%the 100 MHz value of Isobe et al. was replaced with an 86 MHz detection (Mills et 
%al. 1960).

One last highlight comes courtesy of the well-known jet in M87.  The X-ray and
optical intensities of the knot HST-1 in the jet ($\sim$60 pc projected from the
nucleus) had been continually increasing since 2000, reaching a maximum in early
2005 ($\times$50 increase; Harris et al.  2006).  New {\it VLBA} observations were
obtained when HST-1 became bright enough in the radio and revealed several
superluminal (1.4--4$c$) ejections associated with the onset of activity. The
recent {\it HESS} report of a TeV flare (Aharonian et al.  2006) broadly mimicking
the {\it VLA/HST/Chandra} lightcurves of HST-1 give us unique insight into flares
in more distant blazars (C.C. Cheung \& D.E. Harris, in preparation).  Monitoring
is being continued and we expect to extend this program into the {\it GLAST} era.

\section{Compton X-ray to Gamma-ray Emission in Radio Lobes\label{sec-two}}

Since the CMB is ubiquitous, IC/CMB losses are mandatory in synchrotron sources. 
This emission is most prominent in regions of low magnetic field like the extended
lobes of radio galaxies, and many such sources of IC/CMB X-rays are now known
(e.g., Croston et al. 2005; Kataoka \& Stawarz 2005). Perhaps one of the best
examples of IC/CMB X-rays is in the nearby radio galaxy Fornax A (Feigelson et al.
1995) which we further discuss as an exemplary example. 

Fornax A is a bright nearby radio galaxy and integrated radio measurements extend
down to $\sim$30 MHz (Isobe et al.  2006). It is detected out to $\sim$90 GHz by
{\it WMAP} (Fig.~\ref{fig-3}) with a steep cm-wave spectrum,
$F_{\nu}\propto\nu^{-1.5~\pm~0.09}$ (Bennett et al. 2003).  Ignoring kinematic
factors, we can relate the synchrotron to IC/CMB spectra by $\nu_{\rm synch}[{\rm
MHz}] \sim 2 E_{\rm IC}[{\rm keV}] B_{\rm \mu G}/(1+z)$, and figure~\ref{fig-3}
shows the predicted IC component for Fornax A.  This assumes an average
$B$=1.5$\mu$G in the lobes to match the X-ray detections (Feigelson et al.  1995;
Isobe et al. 2006).  The expectation is that the electrons producing radio emission
in the observed $\sim$10's-100 GHz range will produce a hard X-ray/soft
$\gamma$-ray emission signal. Sub-mm observations can show the extent of this
extrapolation in the {\it GLAST} sensitivity range. Fornax A is quite extended in
the sky, so if it is detected by {\it GLAST}, the contributions from the two lobes
will be separable.

\section{Observational Prospects}

Detailed resolved spectral studies utilizing {\it Chandra}, {\it Spitzer}, and {\it
Hubble}, are currently limited mainly to the brightest jets and hotspots.  In the
future, {\it JWST} may detect fainter examples and {\it ALMA} will ``catch'' those
with spectral cutoffs in the $\sim$10's to $\sim$1000 GHz range to better define
their synchrotron continua.  Another course of study enabled by {\it ALMA} is to
search for the signature of ``bulk-Compton'' emission from a relativistic stream of
cold particles (i.e., $<$$\gamma$$>$$\sim$1) scattering the CMB.  In Uchiyama et
al. (2005), our non-detection of this ``bump'' in the mid-IR argues against a pure
e$^+$e$^-$ jet in quasar PKS~0637 if $\Gamma\sim$10--15 as required by an IC/CMB
origin for the X-rays. If instead, jets like this are only mildly relativistic
($\Gamma\sim$2--3), the bulk Compton bump then peaks at $\nu\sim\Gamma^{2}\nu_{\rm
cmb}\sim 160\Gamma^{2}(1+z)$ GHz which is near the {\it ALMA} sensitivity range;
such observations can then test emission models. 

The next generation of low radio frequency interferometers like the {\it Long
Wavelength Array} ({\it LWA}) will provide radio maps with unprecedented
sensitivity and resolution at 10's--100's MHz.  These observations will measure
synchrotron emission from the low-energies electrons which are responsible for IC
emission in the UV/X-ray bands (see Harris 2005 for such applications).  In the
example of Fornax A (\S\ref{sec-two}), the first prototype antenna tiles of the
{\it Mileura Widefield Array} ({\it MWA}) Low-Frequency Demonstrator already show
the double-structure at $\sim$100 MHz (Bowman et al. 2006) that is known at
higher-frequencies.  We can look forward to detailed maps of the various components
of even fainter, more distant radio sources as construction of the arrays
progresses.  If the emission extends into the cm--mm range in regions of relatively
low magnetic field ($\sim$few$\mu$G), it will signal IC/CMB emission in the hard
X-ray/soft $\gamma$-ray bands as postulated in the case of Fornax A;  some of the
brightest examples should be detectable and image-able with {\it GLAST}.

%\acknowledgements Thanks to my many collaborators for their work cited here and my 
%apologies for the inexhaustive citations to their and other work.

\end{document}